\documentclass{appolb}
\usepackage[dvips]{graphicx}% For importing graphics: .ps, .jpg, .bmp, .pcx, &
\usepackage{amsmath}
\usepackage{amsfonts}
\usepackage{amssymb}
\usepackage{longtable} % For tables that are too long for 1 page.
\usepackage{color}

\allowdisplaybreaks[4] % allows page breaks in the middle of an equation
\newcommand{\df}{\ {\overset {\rm def} =}\ }

\newcommand{\dril}[2]{{{\rm d} {#1}} / {{\rm d} {#2}}}
\newcommand{\pdril}[2]{{\partial {#1}} / {\partial {#2}}}

\newcommand{\llim}[1] {\ {\underset {#1} {\longrightarrow}}\ }

\eqsec  % uncomment this line to get equations numbered by (sec.num)

\begin{document}

\title{The Szekeres metrics with $M < 0$}

\author{Andrzej Krasi\'nski
\address{N. Copernicus Astronomical Centre, Polish Academy of Sciences, \\
Bartycka 18, 00 716 Warszawa, Poland} \\
email: akr@camk.edu.pl}

\maketitle
\begin{abstract}
The evolution equation of the Szekeres metrics allows solutions with the mass
function $M < 0$. They exist in both classes of the Szekeres metrics, have no
Big Bang singularity and no origin. In both classes, the conditions for no shell
crossings ensure that the mass density $\rho$ of the dust source in the Einstein
equations is negative at all times. Thus, these metrics do not qualify as
cosmological models. In the Friedmann limit, the implication $M < 0
\Longrightarrow \rho < 0$ is immediate. In the general Szekeres metrics it
follows by tuning conclusions from different equations.
\end{abstract}

\section{Motivation and summary}\label{intro}

\setcounter{equation}{0}

The signature of the metric will be $(+, -, -, -)$, the derivatives will be
denoted as $\pdril f z = f,_z$. We assume the cosmological constant $\Lambda =
0$.

The Szekeres metrics obey the Einstein equations with a dust source and have in
general no symmetry \cite{BSTo1977}. They were found by P. Szekeres in 1975
\cite{Szek1975,Szek1975b} beginning with the following Ansatz
\begin{equation}\label{1.1}
{\rm d} s^2 = {\rm d} t^2 - {\rm e}^{2 \alpha} {\rm d} z^2  - {\rm e}^{2 \beta}
\left({\rm d} x^2 + {\rm d} y^2\right),
\end{equation}
where $\alpha(t, z, x, y)$ and $\beta(t, z, x, y)$ are to be determined by
solving the Einstein equations, and the velocity field of the dust is
$u^{\alpha} = {\delta^{\alpha}}_0$, $(x^0, \dots, x^3) = (t, z, x, y)$. For a
derivation of the solutions of the Einstein equations see Ref. \cite{PlKr2024}.
They split into class I, where $\beta,_z \neq 0$ and class II, where $\beta,_ z
= 0$. Both classes contain the Friedmann metrics \cite{Frie1922,Frie1924} as
limits and class I found numerous applications in theoretical cosmology
\cite{Kras1997}.

The metric of class I (in the parametrisation of Hellaby \cite{Hell1996}) is
\begin{eqnarray}
&& {\rm d} s^2 = {\rm d} t^2 - \frac {\left(\Phi,_z - \Phi {\cal E},_z/{\cal
E}\right)^2} {\varepsilon - k}\ {\rm d} z^2 - \frac {\Phi^2} {{\cal E}^2}
\left({\rm d} x^2 + {\rm d} y^2\right), \label{1.2} \\
&& {\cal E} \df \frac S 2 \left[\left(\frac {x - P} S\right)^2 + \left(\frac {y
- Q} S\right)^2 + \varepsilon\right], \label{1.3}
\end{eqnarray}
where $k, P, Q$ and $S$ are arbitrary functions of $z$, $\varepsilon = \pm 1$ or
0 and the function $\Phi(t, z)$ must obey
\begin{equation}\label{1.4}
{\Phi,_t}^2 = - k + 2M / \Phi.
\end{equation}
The $M(z)$ is an arbitrary function, and the mass density $\rho$ of the dust is
\begin{equation}\label{1.5}
\kappa c^2 \rho = \frac {2 \left(M,_z - 3 M {\cal E},_z / {\cal E}\right)}
{\Phi^2 \left(\Phi,_z - \Phi {\cal E},_z / {\cal E}\right)}, \qquad \kappa \df 8
\pi G / c^4.
\end{equation}
The value of $\varepsilon$ determines the type of {\it quasi-symmetry} of the
metric: with $\varepsilon = +1$ the surfaces of constant $t$ and $z$ in
(\ref{1.2}) -- (\ref{1.3}) are nonconcentric spheres, with $\varepsilon = -1$
they have constant negative curvature, and with $\varepsilon = 0$ they are
intrinsically flat (although it is questionable whether they can be interpreted
as Euclidean planes, a more plausible interpretation is that they are flat tori
\cite{Kras2008}). The corresponding Szekeres metrics are called quasispherical,
quasihyperbolic and quasiplane, respectively.\footnote{The Hellaby
parametrisation of (\ref{1.2}) -- (\ref{1.3}) obscures the fact (which is
evident in the original Szekeres parametrisation) that {\it the same} Szekeres
spacetime may be quasispherical in one region and quasihyperbolic elsewhere, the
boundary between these regions being quasiplane, see Ref. \cite{HeKr2008} for an
illustration.}

The Friedmann metrics result from (\ref{1.2}) -- (\ref{1.4}) when $k = k_0 z^2$,
$M = M_0 z^3$ and $\Phi = z R(t)$ ($k_0$ and $M_0$ being arbitrary constants);
then $R$ obeys an equation identical in form to (\ref{1.4}) with $(k, M, \Phi)$
replaced by $(k_0, M_0, R)$. In the Friedmann limit, ${\cal E},_z = 0$ and
(\ref{1.5}) reduces to $\kappa c^2 \rho  = 6 M_0 / R^3$, so $\rho > 0$
immediately implies $M_0 > 0$. In the general Szekeres case, a solution of
(\ref{1.4}) exists for $M < 0$ and $k < 0$, and it is not immediately clear from
(\ref{1.5}) what consequences $M < 0$ has.

In class II, the metric is (\ref{1.1}) with
\begin{eqnarray}
{\rm e}^{\beta} &=& {\Phi(t)} {\rm e}^{\nu}, \qquad {\rm e}^{\alpha} = \Phi(t)\
\sigma(z, x, y) + \lambda(t, z), \label{1.6} \\
\sigma &=& {\rm e}^{\nu} \left[\frac 1 2 U(z) \left(x^2 + y^2\right) + V_1(z) x
+ V_2(z) y + 2 W(z)\right], \label{1.7} \\
{\rm e}^{- \nu} &\df& 1 + \frac 1 4 k \left(x^2 + y^2\right), \label{1.8}
\end{eqnarray}
where $U, V_1, V_2$ and $W$ are arbitrary functions of $z$. The function
$\Phi(t)$ obeys (\ref{1.4}), but here $k$ and $M$ are constants. We define
\begin{equation}\label{1.9}
- k + 2M / \Phi \df {\cal U}(\Phi).
\end{equation}
The function $\lambda(t, z)$ obeys
\begin{equation}
\lambda,_t \Phi \Phi,_t + \lambda M / \Phi = (U + kW)\Phi + X(z), \label{1.10}
\end{equation}
where $X(z)$ is an arbitrary function. (Equation (\ref{1.10}) is valid only when
$\Phi,_t \neq 0$; this will be assumed.) We will consider only the case $k < 0$;
then $\lambda$ and $W$ can be reparametrised so that the new $\widetilde{W}$
obeys $U + k \widetilde{W} = 0$ \cite{PlKr2024}. Then, omitting the tildes
\begin{equation}\label{1.11}
\lambda,_t \Phi \Phi,_t + \lambda M / \Phi = X(z),
\end{equation}
The solution of (\ref{1.11}) can be written as
\begin{equation}\label{1.12}
\lambda = \sqrt{\cal U} \left[\int \frac {X(z)} {\Phi {\cal U}^{3/2}} {\rm d}
\Phi + Y(z)\right].
\end{equation}
The integral is elementary (see further). The mass density $\rho$ is here
\begin{equation}\label{1.13}
\kappa c^2 \rho = \frac {2X + 6M \sigma} {\Phi^2\ {\rm e}^{\alpha}}.
\end{equation}

Class II metrics can be obtained from class I by a limiting transition invented
by Hellaby \cite{Hell1996b}, see Ref. \cite{PlKr2024} for a derivation.

Although the Szekeres metrics have been investigated in numerous papers with
several applications in view (see Refs. \cite{BCKr2010,BKHC2010} for overviews),
the solutions of (\ref{1.4}) with $M < 0$ have not been looked at up to now, so
their geometrical and physical interpretation is unknown. The aim of the present
paper is to fill this gap.

In Sec. \ref{solns}, the solutions corresponding to $M < 0$ and $k < 0$ are
displayed. In Sec. \ref{sec4clI}, the conditions for avoiding shell crossings in
class I metrics are investigated. It is proved that shell crossings can be
avoided, but the conditions on the arbitrary functions that guarantee the
avoidance imply that the mass density of the dust must be negative at all times.

In Sec. \ref{sec5clII}, the no-shell-crossing conditions are investigated for
class II metrics. Here, if $\rho > 0$ is fulfilled at the instant of maximum
compression, $t = t_B$, then $\rho \to \infty$ at a finite time $t_0$ and $\rho
< 0$ at $t \to \infty$. But the parameters of the metric can be chosen so that
$\rho < 0$ and $|\rho| < \infty$ everywhere.

Section \ref{sumconc} is a summary of the results. The properties listed above
exclude the Szekeres metrics with $M < 0$ as models of any physical object. This
paper is meant to fill a formal gap in the study of the Szekeres metrics. Its
result can be stated as the following theorem:

In the whole family of the Szekeres metrics the inequality $M < 0$ combined with
the conditions of no shell crossings implies that the mass density of the dust
source in the Einstein equations is everywhere negative.

\section{The solutions of (\ref{1.4}) with $M < 0$}\label{solns}

\setcounter{equation}{0}

With $M < 0$, Eq. (\ref{1.4}) has solutions only when $k < 0$. In both classes
of the Szekeres metrics we denote
\begin{equation}\label{2.1}
M = - {\cal M}, \qquad k = - K,
\end{equation}
so that ${\cal M} > 0$ and $K > 0$. Then (\ref{1.4}) becomes
\begin{equation}\label{2.2}
{\Phi,_t}^2 = K - 2 {\cal M} / \Phi.
\end{equation}
The solution of (\ref{2.2}) is
\begin{eqnarray}
\Phi &=& ({\cal M} / K)\ (\cosh \eta + 1), \label{2.3} \\
t - t_B &=& ({\cal M} / K^{3/2})\ (\sinh \eta + \eta), \label{2.4}
\end{eqnarray}
where $\eta$ is the parameter, $t_B$ is an arbitrary function of $z$ in class I
and an arbitrary constant in class II. The $\Phi$ of (\ref{2.3}) does not vanish
at any $\eta \in (- \infty, + \infty)$ and at any $z$ (because ${\cal M} > 0$),
so there is no Big Bang / Big Crunch and no origin. It has the lower bound $2
{\cal M} / K$ at $\eta = 0 \Longleftrightarrow t = t_B$. Eq. (\ref{2.2}) implies
$\Phi,_{t t} = {\cal M} / \Phi^2 > 0$, i.e. the dust expands with acceleration
for $t > t_B$ / contracts with deceleration for $t < t_B$.

Equation (\ref{1.5}) now becomes
\begin{equation}\label{2.5}
\kappa c^2 \rho = \frac {2 \left(3 {\cal M} {\cal E},_z / {\cal E} - {\cal
M},_z\right)} {\Phi^2 \left(\Phi,_z - \Phi {\cal E},_z / {\cal E}\right)}.
\end{equation}

{}From (\ref{2.3}) -- (\ref{2.4}) we find
\begin{equation}\label{2.6}
\Phi,_t = \frac {\sqrt{K} \sinh \eta} {\cosh \eta + 1}.
\end{equation}

{}From this point on, the two classes have to be considered separately.

\section{{\bf Conditions for no shell crossings in class I
($\beta,_z \neq 0$) metrics}}\label{sec4clI}

\setcounter{equation}{0}

Differentiating (\ref{2.3}) -- (\ref{2.4}) by $z$ and using (\ref{2.6}) we find
the equation identical in form to (20.71) in Ref. \cite{PlKr2024}:
\begin{equation}\label{3.1}
\Phi,_z = \left(\frac {{\cal M},_z} {\cal M} - \frac {K,_z} K\right) \Phi +
\left[\left(\frac {3 K,_z} {2K} - \frac {{\cal M},_z} {\cal M}\right) \left(t -
t_B\right) - t_{B,z}\right] \Phi,_t.
\end{equation}
For $\rho$ to be finite, the denominator of (\ref{2.5}) may vanish only together
with the numerator. If $\Phi,_z -\Phi {\cal E},_z / {\cal E} = 0$ anywhere else,
then this is a shell crossing singularity. Thus, to ensure $|\rho| < \infty$ we
must impose the condition
\begin{eqnarray}
{\rm either} && \Phi,_z - \Phi {\cal E},_z / {\cal E} > 0, \label{3.2} \\
{\rm or} && \Phi,_z -\Phi {\cal E},_z / {\cal E} < 0. \label{3.3}
\end{eqnarray}
Now the three types of quasisymmetry must be considered separately.

\bigskip

\centerline{\it 3.1. The quasispherical case}

\bigskip

In this case the set ${\cal E},_z = 0$ must exist and is a great circle of every
$(x, y)$ sphere, while the extrema of ${\cal E},_z / {\cal E}$ are at the poles,
where \cite{HeKr2002}
\begin{equation}\label{3.4}
\left({\cal E},_z / {\cal E}\right)_{\rm extreme} = \pm \frac 1 S\
\sqrt{{S,_z}^2 + {P,_z}^2 + {Q,_z}^2}, \qquad S > 0.
\end{equation}
Let us investigate (\ref{3.2}) first. It will be fulfilled everywhere when
\begin{equation}\label{3.5}
\Phi,_z / \Phi > \left({\cal E},_z / {\cal E}\right)_{\rm max} = + \frac 1 S\
\sqrt{{S,_z}^2 + {P,_z}^2 + {Q,_z}^2}.
\end{equation}
Using (\ref{2.3}) -- (\ref{2.4}), (\ref{2.6}) and (\ref{3.1}), the above can be
written as
\begin{equation}\label{3.6}
\frac {{\cal M},_z} {\cal M} - \frac {K,_z} K + \left(\frac {3 K,_z} {2K} -
\frac {{\cal M},_z} {\cal M}\right) F(\eta) - t_{B,z} \frac {K^{3/2}} {\cal M}
G(\eta) > \left({\cal E},_z / {\cal E}\right)_{\rm max},
\end{equation}
where
\begin{equation}\label{3.7}
F(\eta) \df \frac {\sinh \eta (\sinh \eta + \eta)} {(\cosh \eta + 1)^2}, \qquad
G(\eta) \df \frac {\sinh \eta} {(\cosh \eta + 1)^2}.
\end{equation}
These functions have the following properties
\begin{eqnarray}\label{3.8}
&& F(\eta) > 0\ {\rm for}\ \eta \neq 0, \qquad F(0) = 0, \qquad \lim_{\eta \to
\pm \infty} F(\eta) = 1, \nonumber \\
&& {\rm sign}\ (G) = {\rm sign}\ (\eta), \qquad G(0) = 0, \qquad \lim_{\eta \to
\pm \infty} G(\eta) = 0.
\end{eqnarray}
We want to ensure that (\ref{3.6}) holds at all $\eta$. It will hold at $\eta =
0$ when
\begin{equation}\label{3.9}
\frac {K,_z} K < \frac {{\cal M},_z} {\cal M} - \left(\frac {{\cal E},_z} {\cal
E}\right)_{\rm max},
\end{equation}
and will hold at $\eta \to \pm \infty$ when
\begin{equation}\label{3.10}
K,_z / K > 2 \left({\cal E},_z/{\cal E}\right)_{\rm max}.
\end{equation}
A necessary condition for (\ref{3.9}) and (\ref{3.10}) to hold simultaneously is
\begin{equation}\label{3.11}
{\cal M},_z / {\cal M} > 3 \left({\cal E},_z/{\cal E}\right)_{\rm max}.
\end{equation}
When (\ref{3.5}) holds for all $\eta$, (\ref{2.5}) with (\ref{3.11}) imply that
$\rho < 0$ for all $\eta$. Consequently, this is not a model of the observed
Universe. Nevertheless, we will now show that there exists a choice of ${\cal
M}, K$ and $t_B$ that ensures (\ref{3.6}) to hold at all $\eta$ (which means
that $\rho$ is finite at all times).

Let (\ref{3.11}) hold and let us choose $K$ so that (\ref{3.9}) holds and
\begin{equation}\label{3.12}
\frac {3 K,_z} {2 K} - \frac {{\cal M},_z} {\cal M} > 0.
\end{equation}
It can be verified that $K,_z / K$ and ${\cal M},_z / {\cal M}$ obeying
(\ref{3.9}), (\ref{3.11}) and (\ref{3.12}) do exist, for example by visualising
$\left({\cal E},_z/{\cal E}\right)_{\rm max}$ as a unit of length. Then the
expression in the first line of (\ref{3.6}) alone is greater than $\left({\cal
E},_z/{\cal E}\right)_{\rm max}$ in consequence of (\ref{3.9}) and (\ref{3.12}).

The $G(\eta)$ of (\ref{3.7}) has the further property
\begin{equation}\label{3.13}
- \sqrt{3} / 9 \leq G(\eta) \leq \sqrt{3} / 9.
\end{equation}
Thus, the $G(\eta)$ term in (\ref{3.6}) will enhance the inequality at $\eta >
0$ if
\begin{equation}\label{3.14}
t_{B,z} < 0.
\end{equation}
Eq. (\ref{3.6}) will hold at $\eta < 0$ if it holds at the minimum of $G(\eta)$,
which occurs at such $\eta_0 < 0$ for which $\cosh \eta_0 = 2$ ($\Longrightarrow
\sinh \eta_0 = -\sqrt{3}$), so we require
\begin{eqnarray}\label{3.15}
\frac {{\cal M},_z} {\cal M} - \frac {K,_z} K &+& \left(\frac {3 K,_z} {2K} -
\frac {{\cal M},_z} {\cal M}\right)\ \frac {\sqrt{3} (\sqrt{3} + \eta_0)}
9  \nonumber \\
&-& \left(- t_{B,z}\right) \frac {K^{3/2}} {\cal M}\ \frac {\sqrt{3}} 9 >
\left({\cal E},_z / {\cal E}\right)_{\rm max}.
\end{eqnarray}
which is equivalent to
\begin{eqnarray}\label{3.16}
0 < -t_{B,z} &<& \frac {9 {\cal M}} {\sqrt{3} K^{3/2}}\ \left[\frac {{\cal
M},_z} {\cal M} - \frac {K,_z} K - \left(\frac {{\cal E},_z} {\cal
E}\right)_{\rm max}\right] \nonumber \\
&+& \frac {\cal M} {K^{3/2}}\ \left(\frac {3 K,_z} {2K} - \frac {{\cal M},_z}
{\cal M}\right)\  (\sqrt{3} + \eta_0).
\end{eqnarray}
The expression in the second line is positive in consequence of (\ref{3.9}),
(\ref{3.12}) and of $\sqrt{3} \approx 1.732, \eta_0 \approx - 1.317$.

The final verdict in the (\ref{3.2}) case is thus: if $K$ and ${\cal M}$ obey
(\ref{3.9}) and (\ref{3.12}) and $t_{B,z}$ obeys (\ref{3.16}), then the metric
(\ref{1.2}) with $\varepsilon = +1$ is nowhere singular, but defines a dust
distribution with $\rho < 0$ for all $\eta$.

Now let us verify whether (\ref{3.3}) can co-exist with $\rho > 0$. By the same
reasoning as before we obtain instead of (\ref{3.6})
\begin{eqnarray}\label{3.17}
&& \frac {{\cal M},_z} {\cal M} - \frac {K,_z} K + \left(\frac {3 K,_z} {2K} -
\frac {{\cal M},_z} {\cal M}\right) F(\eta) - t_{B,z} \frac {K^{3/2}} {\cal M}\
G(\eta) < \left({\cal E},_z / {\cal E}\right)_{\rm min} \nonumber \\
&& = - \left({\cal E},_z / {\cal E}\right)_{\rm max}.
\end{eqnarray}
Requiring that (\ref{3.17}) holds at $\eta = 0$ we obtain
\begin{equation}\label{3.18}
\frac {K,_z} K > \frac {{\cal M},_z} {\cal M} + \left(\frac {{\cal E},_z} {\cal
E}\right)_{\rm max}.
\end{equation}
Requiring that (\ref{3.17}) holds at $\eta \to \pm \infty$ we obtain
\begin{equation}\label{3.19}
- K,_z / K > 2 \left({\cal E},_z/{\cal E}\right)_{\rm max}.
\end{equation}
{}From (\ref{3.18}) and (\ref{3.19}) it follows that
\begin{equation}\label{3.20}
- {\cal M},_z / {\cal M} > 3 \left({\cal E},_z/{\cal E}\right)_{\rm max}.
\end{equation}
But for any value of ${\cal E},_z/{\cal E}$ the following is true:
\begin{equation}\label{3.21}
{\cal E},_z/{\cal E} \geq \left({\cal E},_z/{\cal E}\right)_{\rm min} = -
\left({\cal E},_z/{\cal E}\right)_{\rm max}.
\end{equation}
{}From (\ref{3.20}) and (\ref{3.21}) we obtain
\begin{equation}\label{3.22}
3 {\cal E},_z/{\cal E} - {\cal M},_z / {\cal M} > 0,
\end{equation}
and this, together with (\ref{3.3}), implies $\rho < 0$. Thus, even if ${\cal
M}, K$ and $t_B$ can be chosen so that $\Phi,_z / \Phi- {\cal E},_z/{\cal E} <
0$ holds, $\rho < 0$ follows anyway.

\bigskip

\centerline{\it 3.2. The quasihyperbolic case}

\bigskip

This case was worked out in Refs. \cite{HeKr2008} and \cite{KrBo2012} for $M >
0$. When $\varepsilon = -1$, Eq. (\ref{3.4}) is replaced by
\begin{equation}\label{3.23}
\left({\cal E},_z / {\cal E}\right)_{\rm extreme} = - \varepsilon_2 \nu \frac 1
S\ \sqrt{{S,_z}^2 - {P,_z}^2 - {Q,_z}^2},
\end{equation}
where $\varepsilon_2 = \pm 1$ is the sign of $S,_z$ and $\nu = \pm 1$ is the
sign of ${\cal E}$. These extrema exist only if
\begin{equation}\label{3.24}
{S,_z}^2 > {P,_z}^2 + {Q,_z}^2.
\end{equation}
In the opposite case, shell crossings are inevitable with any sign of $M$
\cite{HeKr2008}, so we assume (\ref{3.24}) to hold.

The space of constant $t$ consists of two parts, ${\cal E},_z / {\cal E}$ having
a fixed sign in each part, and nowhere being zero \cite{HeKr2008}. The
$\left({\cal E},_z / {\cal E}\right)_{\rm extreme}$ is a maximum where ${\cal
E},_z / {\cal E} < 0$ and a minimum where ${\cal E},_z / {\cal E} > 0$. (These
two parts are different coordinate coverings of the same space \cite{KrBo2012}.)

Let us first consider (\ref{3.2}), separately for ${\cal E},_z / {\cal E} < 0$
and ${\cal E},_z / {\cal E} > 0$.

\newpage

\centerline{\bf 3.2.1 Case I: ${\cal E},_z / {\cal E} < 0$ with (\ref{3.2})}

\medskip

In this part of the spacetime, the condition for (\ref{3.2}) is
\begin{equation}\label{3.25}
\Phi,_z / \Phi > \left({\cal E},_z / {\cal E}\right)_{\rm max} = - \frac 1 S\
\sqrt{{S,_z}^2 - {P,_z}^2 - {Q,_z}^2} < 0,
\end{equation}
and then the condition for $\rho > 0$ in (\ref{2.5}) is
\begin{equation}\label{3.26}
{\cal M},_z < 3 {\cal M} {\cal E},_z / {\cal E} < 0.
\end{equation}
We can still use (\ref{3.1}) and (\ref{3.6}), $\left({\cal E},_z / {\cal
E}\right)_{\rm max}$ now being (\ref{3.25}) in place of (\ref{3.5}). The
reasoning here proceeds in analogy to the one that follows (\ref{3.8}) and again
leads to (\ref{3.11}), which contradicts (\ref{3.26}). So, $\rho < 0$ also here.

\medskip

\centerline{\bf 3.2.2 Case II: ${\cal E},_z / {\cal E} > 0$ with (\ref{3.2})}

\medskip

In this case, ${\cal E},_z / {\cal E}$ has the positive minimum given by
(\ref{3.23}) with $- \varepsilon_2 \nu = +1$, and is unbounded from above. Thus,
there exists no choice for the functions $K, {\cal M}$ and $t_B$ in (\ref{3.1})
that would guarantee (\ref{3.2}) in the whole region where ${\cal E},_z / {\cal
E} > 0 \Longrightarrow$ this case is empty.

\medskip

We now consider (\ref{3.3}), and again two cases separately:

\medskip

\centerline{\bf 3.2.3 Case III: ${\cal E},_z / {\cal E} < 0$ with (\ref{3.3})}

\medskip

Here, ${\cal E},_z / {\cal E}$ has a negative maximum, but is not bounded from
below. Consequently, it is impossible to fulfil (\ref{3.3}) in this part, and
this case is empty, just like Case II.

\medskip

\centerline{\bf 3.2.4 Case IV: ${\cal E},_z / {\cal E} > 0$ with (\ref{3.3})}

\medskip

Here, ${\cal E},_z / {\cal E}$ has a positive minimum, and the condition for
(\ref{3.3}) is
\begin{equation}\label{3.27}
\Phi,_z / \Phi < \left({\cal E},_z / {\cal E}\right)_{\rm min} = \frac 1 S\
\sqrt{{S,_z}^2 - {P,_z}^2 - {Q,_z}^2} > 0.
\end{equation}
Using (\ref{3.6}) -- (\ref{3.7}) for $\Phi,_z / \Phi$, the reasoning presented
below (\ref{3.8}) leads to
\begin{equation}
{\cal M},_z / {\cal M} < 3 \left({\cal E},_z/{\cal E}\right)_{\rm min}.
\label{3.28}
\end{equation}
This guarantees that $\rho < 0$. So, the conclusion here is the same as in Case
I. The equivalence of the results obtained for ${\cal E},_z / {\cal E} < 0$ and
for ${\cal E},_z / {\cal E} > 0$ confirms that the two parts of each $t = $
constant space are different coordinate coverings of the same object
\cite{KrBo2012}.

\newpage

\centerline{\it 3.3. The quasiplane case}

\bigskip

This case was worked out in Refs. \cite{HeKr2008} and \cite{Kras2008} for $M >
0$. When $\varepsilon = 0$, the metric (\ref{1.2}) -- (\ref{1.3}) and Eqs.
(\ref{1.4}) -- (\ref{1.5}) do not change in form when the functions in them are
redefined as follows:
\begin{equation}\label{3.29}
(\Phi, {\cal E}) = \frac 1 S\ (\widetilde{\Phi}, \widetilde{\cal E}), \qquad K =
\widetilde{K} / S^2, \qquad {\cal M} = \widetilde{\cal M} / S^3.
\end{equation}
The result is the same as if $S = 1$, and we shall assume this.

As shown in Ref. \cite{Kras2008}, a surface of constant $t$ and $z$ in the
quasiplane Szekeres model can be interpreted in two ways:

1. As an infinite plane.

2. As a torus-like closed surface that is created by identifying opposite edges
of a flat basic square.\footnote{Actually, the basic square consists of four
smaller squares whose edges are identified with a twist \cite{Kras2008}. The
twist is necessary to fulfil the matching conditions.}

With the first interpretation, shell crossings are inevitable
\cite{HeKr2008,Kras2008}, so we shall not consider it here.

With the second interpretation, to avoid shell crossings the
inequality\footnote{Tildes are dropped for better readability. The $\Phi$ below
is the $\widetilde{\Phi}$ of (\ref{3.29}).}
\begin{equation}\label{3.30}
\frac {\Phi,_z} {\Phi} \geq \frac {\pi} {\sqrt{2}}\ \sqrt{{P,_z}^2 + {Q,_z}^2}
\end{equation}
must be fulfilled at all $t$ and $z$ \cite{Kras2008}; then the set where
$\Phi,_z / \Phi - {\cal E},_z / {\cal E} = 0$ lies outside the basic square,
i.e. is not part of the spacetime. The calculation here is analogous to the one
that we applied to (\ref{3.5}) -- (\ref{3.6}), Eq. (\ref{3.11}) follows again,
and $\rho < 0$ is implied also here.

\section{Conditions for no shell crossings in Class II ($\beta,_z = 0$)
metrics}\label{sec5clII}

\setcounter{equation}{0}

Unlike in class I, the sign of $k$ determines the type of quasi-symmetry, and
with $k < 0$ the metric is quasihyperbolic. Equations (\ref{2.2}) -- (\ref{2.4})
still hold, but now $K > 0$ and ${\cal M} > 0$ are constants. The integral in
(\ref{1.12}) reads (recall: ${\cal U} = K - 2 {\cal M} / \Phi$):
\begin{equation}\label{4.1}
\int \frac {{\rm d} \Phi} {\Phi {\cal U}^{3/2}} = - \frac 2 {K \sqrt{\cal U}} +
\frac 1 {K^{3/2}}\ \ln \left|\frac {\sqrt{K} + \sqrt{\cal U}} {\sqrt{K} -
\sqrt{\cal U}}\right|.
\end{equation}
Note that $0 \leq {\cal U} \leq K$ from (\ref{1.9}), so $(\sqrt{K} + \sqrt{\cal
U}) / (\sqrt{K} - \sqrt{\cal U}) \geq 1$, and consequently the ln-term in
(\ref{4.1}) is nonnegative at all $t$. In fact, the whole right-hand side of
(\ref{4.1}) is nonnegative.\footnote{This is an easy exercise for the reader.
Begin by defining $x \df \sqrt{K / {\cal U}} \geq 1$.}

The explicit formula for $\lambda$ becomes now
\begin{equation}\label{4.2}
\lambda = X(z) \left[- \frac 2 K + \frac {\sqrt{\cal U}} {K^{3/2}}\ \ln
\left|\frac {\sqrt{K} + \sqrt{\cal U}} {\sqrt{K} - \sqrt{\cal U}}\right|\right]
+ Y(z) \sqrt{\cal U}.
\end{equation}
The formula for mass density (\ref{1.13}) is here
\begin{equation}\label{4.3}
\kappa c^2 \rho = \frac {2X - 6 {\cal M} \sigma} {\Phi^2\ {\rm e}^{\alpha}}.
\end{equation}
and (\ref{1.8}) is
\begin{equation}\label{4.4}
{\rm e}^{- \nu} = 1 - \frac 1 4\ K \left(x^2 + y^2\right).
\end{equation}
Seemingly, this spacetime consists of two disjoint regions, $x^2 + y^2 > 4 / K$
and $x^2 + y^2 < 4 / K$. But the argument used in Ref. \cite{KrBo2012} for the
quasi\-hyperbolic class I metrics applies also here: the two regions are
different coordinate coverings of the same spacetime, so it suffices to consider
one of them.

We verify whether it is possible to choose the arbitrary functions so that
$|\rho| < \infty$ and $\rho > 0$ everywhere. We begin by assuming
\begin{equation}\label{4.5}
X > 3 {\cal M} \sigma
\end{equation}
and ${\rm e}^{\alpha} > 0$ in (\ref{4.3}). From (\ref{1.6}) and (\ref{4.2}) we
have
\begin{equation}\label{4.6}
{\rm e}^{\alpha} = \Phi \sigma - \frac {2X} K + \frac {X \sqrt{\cal U}}
{K^{3/2}}\ \ln \left|\frac {\sqrt{K} + \sqrt{\cal U}} {\sqrt{K} - \sqrt{\cal
U}}\right| + Y \sqrt{\cal U}.
\end{equation}
Taking this at $t = t_B$, where $\Phi = 2 {\cal M} / K$ and ${\cal U} = 0$, we
obtain from ${\rm e}^{\alpha} > 0$
\begin{equation}\label{4.7}
X < {\cal M} \sigma.
\end{equation}
This can hold simultaneously with (\ref{4.5}) only when $\sigma < 0$, so $X <
0$.

To see what happens as $t \to \infty$, where $\Phi \to \infty$, let us note that
\begin{equation}\label{4.8}
\lim_{t \to \infty} \left[\frac 1 {\Phi}\ \ln \left|\frac {\sqrt{K} + \sqrt{\cal
U}} {\sqrt{K} - \sqrt{\cal U}}\right|\right] \equiv \lim_{t \to \infty}
\left[\frac 2 {\Phi}\ \ln \left|\frac {\Phi} {2 {\cal M}}\ \left(\sqrt{\cal U} +
\sqrt{K}\right)\right|\right] = 0,
\end{equation}
so, at large $t$, $|\Phi \sigma|$  dominates over the other terms in
(\ref{4.6}). With $\sigma < 0$, this means $\lim_{t \to \infty} {\rm e}^{\alpha}
= - \infty$, i.e. $\rho$ becomes negative in the infinite future. But if ${\rm
e}^{\alpha} > 0$ at $t = t_B$ and ${\rm e}^{\alpha} < 0$ at $t \to \infty$, then
${\rm e}^{\alpha} = 0$ at some $t_0 > t_B$.\footnote{From (\ref{4.6}) $\dril
{{\rm e}^{\alpha}} t$ is well defined for all $t_B < t < \infty$, so ${\rm
e}^{\alpha}$ is continuous.} Then (\ref{4.3}) shows that $\rho \llim{t \to t_0}
\infty$, so a singularity in $\rho$ is inevitable.

Let us now assume instead of (\ref{4.5})
\begin{equation}\label{4.9}
X < 3 {\cal M} \sigma.
\end{equation}
Is this compatible with ${\rm e}^{\alpha} < 0$? Taking ${\rm e}^{\alpha} < 0$
and (\ref{4.6}) at $t = t_B$ we obtain
\begin{equation}\label{4.10}
X > {\cal M} \sigma,
\end{equation}
Equations (\ref{4.9}) and (\ref{4.10}) are consistent only when $\sigma > 0$.
But then, from (\ref{4.8}), $\Phi \sigma$ dominates over the other terms in
(\ref{4.6}) as $t \to \infty$, so ${\rm e}^{\alpha} \llim{t \to \infty} +
\infty$, and again ${\rm e}^{\alpha}$ must go through zero somewhere between $t
= t_B$ and $t \to \infty$.

So, if $\rho > 0$ at the minimum of $\Phi$, then $\rho$ becomes infinite at some
$t = t_0
> t_B$ and negative at $t \to \infty$.

Let us now see whether $\rho$ can be nonsingular when $\rho < 0$. Assuming ${\rm
e}^{\alpha} < 0$ and taking ${\rm e}^{\alpha}$ at $t = t_B$ we obtain
(\ref{4.10}) again. At $t \to \infty$ the sign of ${\rm e}^{\alpha}$ is the same
as the sign of $\sigma$. So, to ensure $\rho < 0$ at large $t$ we must have $X >
3 {\cal M} \sigma$ when $\sigma > 0$ and $X > {\cal M} \sigma$ when $\sigma <
0$.

Let us take $\sigma > 0$ for the beginning. The inequality $X > 3 {\cal M}
\sigma$ will hold throughout the whole $(x, y)$ surface when $\sigma$ has an
upper bound $\sigma_{\rm max}$, and $X > 3 {\cal M} \sigma_{\rm max}$. Let us
consider the region where ${\rm e}^{- \nu} > 0$ in (\ref{1.8}):
\begin{equation}\label{4.11}
x^2 + y^2 < 4 / K
\end{equation}
and let us denote
\begin{equation}\label{4.12}
{\cal P} \df \frac 1 2\ U \left(x^2 + y^2\right) + V_1 x + V_2 y + 2U / K
\end{equation}
(recall: we use the parametrisation in which $W = - U / k = U / K$), so
\begin{equation}\label{4.13}
\sigma = \frac {\cal P} {1 - \tfrac 1 4\ K \left(x^2 + y^2\right)} .
\end{equation}
Let the overbar denote the limit at $x^2 + y^2 \nearrow 4 / K$. Then ${\overline
\sigma} = \pm \infty$ and the sign of $\sigma$ is the sign of ${\cal P}$. Note
that ${\overline {\cal P}}$ is finite. If ${\overline {\cal P}} > 0$, then
$\sigma$ has no upper bound. If ${\overline {\cal P}} < 0$, then ${\overline
{\sigma}} = - \infty$, contrary to the assumed $\sigma > 0$. So, $\sigma > 0$
throughout the disk (\ref{4.11}) is impossible.

So now let $\sigma < 0$. Then ${\cal P} < 0$ within the disk (\ref{4.11}), so
${\overline {\cal P}} \leq 0$ (${\overline {\sigma}} = - \infty$ if
${\overline{\cal P}} < 0$) and $X > {\cal M} \sigma$ (this still permits $X <
0$) in (\ref{4.6}). The $\Phi(t)$ is an increasing function at $t > t_B$, so
$\Phi(t) \sigma < \Phi(t_B) \sigma < 0$. Moreover, the co-factor of $X$ in
(\ref{4.6}), which is proportional to the right-hand side of (\ref{4.1}), is
positive for all $t$. Then a sufficient condition for ${\rm e}^{\alpha} < 0$ in
the disk $x^2 + y^2 < 4 / K$ is $X < 0$ and $Y < 0$.

To ensure $\sigma < 0$ in the disk (\ref{4.11}) we must have ${\cal P} < 0$,
even at the maximum of ${\cal P}$. The maximum is at $(x, y)_{\rm max} = - (1 /
U) (V_1, V_2)$, $U < 0$, and the point $(x, y)_{\rm max}$ will be within the
disk (\ref{4.11}) when
\begin{equation}\label{4.14}
{V_1}^2 + {V_2}^2 < 4 U^2 / K,
\end{equation}
and then ${\overline {\cal P}} < 0$, as it should be.

So, the final conclusion is that in class II $\rho > 0$ and $|\rho| < \infty$ is
impossible, but $\rho < 0$ and $|\rho| < \infty$ is achieved by choosing the
constants and functions so that (\ref{4.14}) holds, and
\begin{equation}\label{4.15}
U < 0, \qquad {\cal M} \sigma_{\rm max} < X < 0 \qquad {\rm and} \qquad Y < 0.
\end{equation}

\section{Summary and conclusions}\label{sumconc}

\setcounter{equation}{0}

The evolution equation of the Szekeres metrics, (\ref{1.4}) (which is identical
to the Friedmann equation), has a well-defined solution when the mass function
$M$ is negative (\cite{PlKr2024}, Sec. 20.2.1; then also $k < 0$). In the
Friedmann limit, $M < 0$ immediately implies $\rho < 0$ for the mass density,
which is impossible in any real object. However, in the Szekeres metrics, where
the relation between $M$ and $\rho$ is (\ref{1.5}) in class I and (\ref{1.13})
in class II, the consequences of $M < 0$ are not immediately visible. The aim of
the present paper was to investigate them.

In brief, when $M < 0$, for both classes of the Szekeres metrics the conditions
for no shell-crossing lead to $\rho < 0$ in the whole spacetime. Consequently,
the Szekeres metrics with $M < 0$ cannot describe any physical entity.

Section \ref{intro} presented the motivation for the present paper and the basic
properties of both classes of the Szekeres metrics. In Sec. \ref{solns}, the
explicit solution of the Szekeres evolution equation with $M < 0$ was given.

In Sec. \ref{sec4clI}, the conditions for no shell crossings in class I metrics
were investigated for the quasispherical, quasihyperbolic and quasiplane cases
with $M < 0$ and $k < 0$. In each case, the shell crossing is avoided at the
cost of having $\rho < 0$ throughout the spacetime.

In Sec. \ref{sec5clII}, the reasoning of Sec. \ref{sec4clI} was applied to class
II metrics. Here, with $k < 0$, only the quasihyperbolic metric exists. The
conclusion was that if $\rho > 0$ at the instant of maximum compression, then
necessarily $\rho \to \infty$ at a finite $t = t_0$, so a singularity is
unavoidable. But the parameters of the metric can be chosen so that $|\rho| <
\infty$ and $\rho < 0$ everywhere.

The property $\rho < 0$ excludes the Szekeres metrics with $M < 0$ as models of
any physical object, and $\rho \to \infty$ at a finite $t$ is an even worse
offence. This paper was meant to fill a formal gap in the investigation of the
Szekeres family of metrics: the case $M < 0$ has not been looked at so far.

\end{document}